\documentclass[12pt]{article}

\usepackage{amsmath}
\usepackage{graphicx}
\usepackage{enumerate}
\usepackage{natbib}
\usepackage{url} 

\RequirePackage[OT1]{fontenc}

\RequirePackage{hypernat}
\RequirePackage{latexsym}
\RequirePackage{float,epsfig,multirow,rotating,times}
\RequirePackage{upgreek,wrapfig}
\RequirePackage{comment}
\RequirePackage[english]{babel} 
\RequirePackage{textcomp}

\newcommand{\bgamma}{\boldsymbol{\gamma}}

 \newcommand{\bitem}{\begin{itemize}}
 \newcommand{\eitem}{\end{itemize}}

\usepackage{authblk}
\providecommand{\keywords}[1]{\textbf{\textit{Keywords:}} #1}

\usepackage{threeparttable}

\newcommand*{\affaddr}[1]{#1} 
\newcommand*{\affmark}[1][*]{\textsuperscript{#1}}
\newcommand*{\email}[1]{\texttt{#1}}

\begin{document}

\title{\bf Hierarchical Bayesian integrated modeling of age- and sex-structured 
wildlife population dynamics}

%

\author{%
Sabyasachi Mukhopadhyay\affmark[1,2], Hans-Peter Piepho\affmark[1], Sourabh Bhattacharya\affmark[3], Holly T. Dublin\affmark[4], Joseph O. Ogutu\affmark[1]\\
\affaddr{\affmark[1]University of Hohenheim, Institute of Crop Science, Biostatistics Unit (340C), Fruwirthstrasse 23, \\ 70599 Stuttgart Germany}\\
\affaddr{\affmark[2]Indian Institute of Management, Udaipur, India}\\
\affaddr{\affmark[3]ISRU, Indian Statistical Institute, \\ 203, B.T. Road, Kolkata-108, India}\\
\affaddr{\affmark[4]IUCN ESARO, Wasaa Conservation Centre, \\ P.O. Box 68200, Nairobi, Kenya, 00200}\\
Email for correspondence: \email{mukhopad@uni-hohenheim.de, sabymstat@gmail.com, sabyasachi.mukhopadhyay@iimu.ac.in}
}
\setcounter{Maxaffil}{0}
\renewcommand\Affilfont{\itshape\small}

\maketitle

\begin{abstract} 
Biodiversity of large wild mammals is declining at alarming rates worldwide. It is therefore 
imperative to develop effective population conservation and recovery strategies. Population dynamics models 
can provide insights into processes driving declines of particular populations of a species and their relative
importance. Accordingly, we develop an integrated Bayesian state-space population dynamics model for wildlife populations 
and illustrate it using a topi population inhabiting the Masai Mara Ecosystem in Kenya. The model is general and 
integrates ground demographic survey with aerial survey monitoring data. It incorporates population 
age- and sex-structure and life-history traits and strategies and relates birth rates, age-specific survival rates and sex 
ratios with meteorological covariates, prior population density, environmental seasonality and predation risk. 
It runs on a monthly time step, enabling accurate characterization of reproductive seasonality, 
phenology, synchrony and prolificacy of births, juvenile and adult recruitments. Model performance is evaluated 
using balanced bootstrap sampling and by comparing model predictions with empirical aerial population size 
estimates. The hierarchical Bayesian model is implemented 
using MCMC methods  for parameter estimation, prediction and inference and reproduces several well-known features of the Mara topi population, including striking and 
persistent population decline, seasonality of births, juvenile and adult recruitments. It is general and can be 
readily adapted for other wildlife species and extended to incorporate several additional useful features.
\end{abstract}


\maketitle

\section{{\bf Introduction} \label{sec:intro}} 

Biodiversity is declining worldwide at such an alarming rate that biologists have christened the contemporary
biodiversity loss as the sixth mass extinction 
\citep{McCallum2015,Ceballosetal2017}. Large mammal populations are particularly at risk
in many ecosystems. Across continental Africa, many populations of large mammal 
species are undergoing disturbing declines  
\citep{Craigieetal2010,Chaseetal2016}. In Kenya, for example, large herbivore populations 
declined by 70\% on average between 1977 and 2016 
\citep{Ogutupiepho2011,Ogutupiephoetal2016}.   
It is therefore imperative to advance our understanding of large herbivore
population dynamics as a basis for developing 
species conservation and management and population recovery strategies. A reliable population
dynamics model can help quantify and evaluate the relative importance of multiple
processes driving declines of particular populations of a species.

For populations inhabiting seasonally variable environments and reproducing 
seasonally, such models can help quantify shifts in seasonality, phenology,
synchrony and prolificity of births, juvenile recruitment and sex ratio in response to 
climate and other changes. The models can also be used to estimate 
population trajectories and assess 
likely population responses to conservation and management interventions, 
projected future scenarios of climate change, 
human population growth, socio-economic development, land use and other changes
(\citeauthor{Zhao2019a}, \citeyear{Zhao2019a}; 
\citeauthor{Zhao2019b}, \citeyear{Zhao2019b}). 

Animal population dynamics models often use independent data collected using various
methods, such as ground demographic surveys and aerial surveys. 
Population dynamics models are also increasingly using  information
from multiple sources to make inferences on various features of populations.
Notably, integrated population models are becoming widely used to 
combine different types of data from disparate sources to make joint inferences on animal
population dynamics 
\citep{Trenkeletal2000,Besbeasetal2005,Rhodesetal2011,Maunderetal2015,Mosnieretal2015}. However, thus far, integrated 
state-space population dynamics models for wildlife populations that incorporate realistic life-history traits and 
strategies and fine temporal frames have not been developed.  Yet, such models are crucial for 
understanding and quantifying how life-history traits and strategies shape population responses to trophic 
interactions, natural environmental and anthropogenic stressors.

Here, we develop an integrated population dynamics model for large wild 
herbivores that integrates aerial survey data with fine-resolution ground demographic 
survey data. The model can be used to predict large herbivore population
dynamics and evaluate the relative importance of various factors driving
population dynamics.
It accounts for influences on large herbivore population dynamics
of variation in climatic 
components, notably rainfall, temperature and their interactions; predation, 
density-dependence, population age and sex (adult sex ratio) structure, gestation length, weaning 
period, adult female pregnancy status, adult females available to conceive, females reaching the age of first-time 
conception, birth, juvenile and adult recruitment,
age- and sex-specific survival rates and environmental seasonality. The model runs on a discrete monthly time step to reliably
track temporal variation in female pregnancy status, birth, juvenile and adult
recruitment, age and sex-specific survival rates, adult sex ratio, population size and inter-annual
variation in reproductive seasonality, phenology, synchrony and prolificity of 
births. The model is 
illustrated using the topi ({\it Damaliscus lunatus korrigum}) 
population inhabiting the Masai Mara Ecosystem of south-western Kenya but is general 
and applicable to populations of other large herbivore species. 

Our integrated state-space modelling approach has several attractive and desirable properties. 
(1) It integrates various data types from multiple sources, notably aerial surveys  and demographic
ground counts. (2) It incorporates realistic population life-history traits and  strategies, survival 
rates, sex ratios and trophic interactions, namely predation and competition. (3) It accommodates 
and permits straightforward representation of possibly complex non-linear relationships between 
birth and adult rectruitments, age-specific survival rates, adult sex ratio and multiple covariates. 
(4) It predicts the ecosystem-wide population size and distributes this among the ecosystem's 
subregions, including those covered by the aerial surveys only. (5) The model uses  a monthly time 
step, allowing accurate characterization of reproductive seasonality, phenology, synchrony and 
prolificity of birth and juvenile recruitment. (6) The model allows efficient computation of
posterior distributions of many parameters and uses a flexible Transformation MCMC (TMCMC) 
technique to enhance computational efficiency and accelerate convergence of iterations.  (7) The model
is validated using balanced bootstrap sampling to generate multiple population trajectories and 
establish robustness. (8) It uses the Importance Resampling MCMC (IRMCMC)  technique for the first
time to accelerate MCMC iterations for multiple data sets each of which  involves high computational
costs. (9) Lastly, the model is robust and reliably reproduces well  known features of animal 
population dynamics and can be easily adapted for other species.

The rest of the paper is organized as follows. We describe the data in 
Section \ref{sec:data}. The Bayesian state-space model is described in 
Section \ref{sec:aim_model} along with an evaluation of its performance
using balanced bootstrap samples. In Section \ref{subsec:surv_eqns}, we 
formulate the birth and adult recruitments, age-specific survival and adult sex ratio models, 
prior distribtions and other model components. In Section 
\ref{sec:post_dist_converge_mcmc}, we discuss the convergence
of the MCMC chain and model validation.
In Section \ref{sec:results}, we present results of applying the 
state-space model to the Mara-Serengeti topi population.
Finally, in Section \ref{sec:discussion} we discuss the results and extensions
of the model.

\section{{\bf The data} \label{sec:data}}

We provide brief overviews of the data and the study population in Sections 2.1 to 2.4 and supply additional
details on the vehicle ground sample counts, aerial sample surveys, the 
Mara-Serengeti topi population, rainfall and temperature
in Web Appendices B-E. The full 
data dictionary providing definitions of all the variables and 
transformations used in the model is provided in Web Appendix A.

\subsection{{\bf Ground vehicle age and sex composition sample surveys} \label{subsec:ground_survey_data}}

Vehicle ground age and sex composition sample surveys of seven ungulate species, including topi, 
were carried out monthly in the Maasai Mara National Reserve and its adjoining pastoral lands for 
174 months from July 1989 to December 2003. Approximate ages of topi in each size class
are newborn (Age $<$ 1 month), quarter size class (1 month $\leq$ Age $<$ 6 months), 
half-yearling class (6 months $\leq$ Age $<$ 19 months) and adults (19 months $\leq$ Age). 
Adult topi were not aged but were sexed using the presence, size and shape of horns, dimorphic 
morphology of the external genitalia and other secondary sexual characters. 
During the entire monitoring period 91,582 topi were aged and 78,738 were sexed 
\citep{OgutuPiephoDublin2008a}. The age- and sex-structured topi ground counts are 
provided in Web Appendix A and described in Web Table 1.

\subsection{{\bf Aerial sample surveys} \label{aerial_sample_data}}

These data are independent of the vehicle ground sample age- and sex-structured counts. 
The Directorate of Resource Surveys and Remote Sensing of Kenya (DRSRS) 
monitored wildlife population size and distribution in the Maasai Mara Ecosystem (6665.6 km$^2$) using 
systematic reconnaissance flights from 1977. A total of 75 surveys were carried out in the 
ecosystem from 1977 to 2018 using 662 flights. The total topi population size estimated from the aerial
survey data are provided in Web Appendix A and described in Web Table 2.

\subsection{{\bf The Mara-Serengeti topi population} \label{subsec:mara_serengeti_topi}}

Topi is a resident grazer in the Mara--Serengeti ecosystem. There, births are seasonal,
start in July and peak at the onset of the early 
rains in October-November, whereas conceptions peak at the start of the long rains
in February-March. Births occur in all months but are rare from January to July 
(\citeauthor{SinclairMduma2000}, \citeyear{SinclairMduma2000};
\citeauthor{OgutuPiepho2010}, \citeyear{OgutuPiepho2010}; \citeyear{OgutuPiepho2014},
\citeyear{Ogutuetal2015b}).
The gestation period of 8 months is followed by a lactation 
period of 3 months. Consequently, topi young are weaned after 3--4 months and nursing ceases 
before conceptions. Topi thus take about 11 months from one conception cycle to the next and 
give birth to one young per year. The young go through a hiding stage before following their 
mothers \citep{Estes1991, SkinnerChimimba2005}. Females attain sexual maturity after 
about 18 months. Topi pregnancy rate in Mara-Serengeti is 100\% for adult but
86\% for 2-year-old females \citep{Duncanthesis1975}. 

\subsection{{\bf Rainfall and temperature} \label{subsec:rainfall_temperature}}

In African savannas vegetation production, quantity and quality are controlled by rainfall
(\citeauthor{Deshmukh1984}, \citeyear{Deshmukh1984}; \citeauthor{Bouttonetal1988a},
\citeyear{Bouttonetal1988a}).
Rainfall seasonality generates and sustains seasonality in food availability and quality for large 
herbivores \citep{Bouttonetal1988b}. Accordingly, rainfall indexes food availability and quality for savanna herbivores. Seasonal 
temperature fluctuations additionally 
affect food quality for herbivores by governing the retention period of green plant leaf 
through the dry season. Monthly rainfall was averaged over a network of 
15 monitoring gauges spread over the Mara to account for spatial variation \citep{Mukhopadhyayetal2019}. 
The monthly averages of blended satellite-station maximum and minimum temperatures data for each 
5 km $\times$ 5 km grid cell in the Mara Ecosystem were also extracted from the Chirps 
data \citep{Funketal2015} and used as covariates. The rainfall and temperature components,
the months covered by each component, moving averages, lags and lagged moving averages computed
for each component and used as predictors of birth and recruitments, survival probabilities and adult sex ratio are
described in Web Table 3.

\section{{\bf The integrated population dynamics state-space model} \label{sec:aim_model}}

We construct a general age- and sex-structured  
population dynamics model for large wild herbivores. The model uses the 
number of animals observed in the ground and aerial surveys 
(Sections \ref{subsec:ground_survey_data} and \ref{aerial_sample_data}) 
that are only samples from
the unobserved true population about which we wish to make inferences.
Thus, the population dynamics model entails two parallel but connected 
processes. The first is the
unobserved true population that evolves over time (called state process) and 
the second is the variation in the observed counts over time (called 
observation process). A common approach to 
modelling both processes simultaneously is to use state-space models 
(\citeauthor{Thomasetal2005}, \citeyear{Thomasetal2005};
\citeauthor{Bucklandetal2007}, \citeyear{Bucklandetal2007};
\citeauthor{Newmanetal2009}, \citeyear{Newmanetal2009};
\citeauthor{Newmanetal2014}, \citeyear{Newmanetal2014}; 
\citeauthor{Maunderetal2015}, \citeyear{Maunderetal2015}).

We develop an integrated population dynamics state-space model which 
couples a hypothetical (or latent) mechanistic model of large herbivore population dynamics 
(state process model), with a statistical observation model of aerial survey and ground 
demographic data (observation process model). In the state-space model, the state 
process model predicts the true but unknown future state of the large herbivore 
population given its current state. The observation
model weights the predictions by the likelihood of the data and thus links the 
process model to the observations. Consequently, the model integrates the aerial 
survey monitoring data with the contemporaneous but independent ground 
demographic survey data.

The state-space model involves quantifying birth recruitment and survival rates
of various age and sex classes of the population and sex ratio as functions of climatic 
factors (e.g., rainfall and temperature), intraspecific
competition, population density, predation  and seasonality. Our state-space 
model shares similarities with the general approaches proposed by 
\cite{Bucklandetal2004} and 
\citeauthor{Newmanetal2006} (\citeyear{Newmanetal2006}; \citeyear{Newmanetal2014}) but also has some notable 
differences. In particular, we make different distributional assumptions for
the initial states and the other components of the state and observation 
processes. Most crucially, our approach differs from theirs with respect to
several structural assumptions and our proposal to model transition probabilities of the state process using 
log-linear models in which covariates such as rainfall, temperature and 
population density are used as predictors of birth recruitment and survival
probabilities and adult sex ratio, similar to the Bayesian approach of 
\cite{Brooksetal2004}. Also, unlike \cite{Newmanetal2006}, we illustrate our 
model using a non-migratory species but the model can be easily adapted for migratory species. Lastly, our model incorporates several 
key life-history traits and strategies crucial to understanding large herbivore population 
biology and dynamics and uses rare long-term fine-resolution ground demographic 
and aerial survey monitoring data.

In Sections \ref{sec:des_model} to \ref{subsec:prior_distrib}, we describe the state 
process and observation models and the associated 
notations. In particular, we describe parameters of the state process model and how they link
birth and adult recruitments and age- and sex-specific survival probabilities and adult sex ratio with covariates in Sections
\ref{subsec:notations_cov} to \ref{subsec:deter_surv_eqns}. Accurate estimation of 
birth and adult recruitments and age- and sex-specific survival 
probabilities and sex ratio is therefore a crucial step in developing the state-space model. As a result, 
the model explicitly allows for the dependence of birth and adult recruitments and age- and sex-specific 
survival probabilities and sex ratio on food availability and quality, density-dependent intraspecific 
competition for food and large carnivore predation. Influences of these factors are indexed by
past rainfall, minimum and maximum temperatures and their interactions, prior total population
size and environmental seasonality.

The relatively large number of parameters considerably complicates their estimation using 
classical techniques. We overcome this difficulty using a flexible Bayesian state-space model 
and present forms of the prior and full conditional distributions in Sections \ref{subsec:prior_distrib}
and \ref{sec:post_dist_converge_mcmc} and in Web Appendix G.

\subsection{{\bf The age and sex structured state-space model--process model} \label{sec:des_model}}
 
To construct the age- and sex-structured state-space model we first introduce notations for the
different topi age and sex classes at time $t$ of the observation process as follows.
We assume throughout that all births or recruitments occur at the end of each month.
$new(t)$ = observed number of newborns, $q(t)$ = observed number of quarter-sized animals,
$h(t)$ = observed number of half-yearlings, $f(t)$ = observed number of adult females,
$m(t)$ = observed number of adult males. The state process involves the same age and sex classes.
The true but unknown numbers of 
animals in each age and sex class are denoted by 
$New(t)$ = actual number of newborns, $Q(t)$ = actual number of quarters,
$H(t)$ = actual number of half-yearlings, $F(t)$ = actual number of adult females,
$AM(t)$ = actual number of adult males.

\subsection{{\bf Assigning topi to actual ages in months} \label{subsec:split_age_class}}

Since the exact age of topi is hard to determine through visual field observation,
animals were only assigned to age and sex classes. However, the probability of survival likely 
varies with age and other temporally varying covariates, such as rainfall. 
For example, a newborn topi must survive the 
first month of its life to join the quarter-size class. Likewise, a 
quarter-size topi must survive through 5 consecutive months before 
graduating to the half-yearling class. 
So, we assign topi in the quarter and half-yearling age classes to actual ages
in months as follows.
$Q(t, k)$ = Number of individuals in the quarter-size class with actual ages lying 
between $k-1$ and $k$ ($k-1$ included but $k$ excluded), $k = 2, \ldots, 6$ months;
$H(t, k)$ = Number of individuals in the half-yearling class with actual ages lying between $k-1$ and 
$k$ ($k-1$ included but $k$ excluded), $k = 7, \ldots, 19$ months.
Note that $Q(t) = \sum_{k=1}^{6} Q(t, k)$ and $H(t) = \sum_{k=1}^{19}H(t, k)$.

For adult females, tracking the reproductive cycle is essential for understanding population 
dynamics. Young topi start reproducing at about 19 months of age. The topi reproductive cycle spans
11 months, including 8 months for gestation and 3 months for lactation. We assume that a female 
cannot conceive during this 11-month period. If pregancy is prematurely terminated, however, 
then a fresh conception may occur within the 11-month period. But we do not have data to
estimate the probability that a pregnant female topi fails to carry pregnancy to term and so do
not consider it in the model. We track the pregnancy status of adult females in each of the the
11 months spanned by the reproductive cycle as follows:  
$AF(t, \ell)$ = Number of adult females at time $t$, that gave birth
exactly $\ell$ months ago, $\ell = 1, \ldots, 11$;
$AF(t, 12)$ = Number of adult females at time $t$, that gave birth
at least 12 months ago. 

\begin{figure}
\centering
\includegraphics[width=0.7\textwidth]{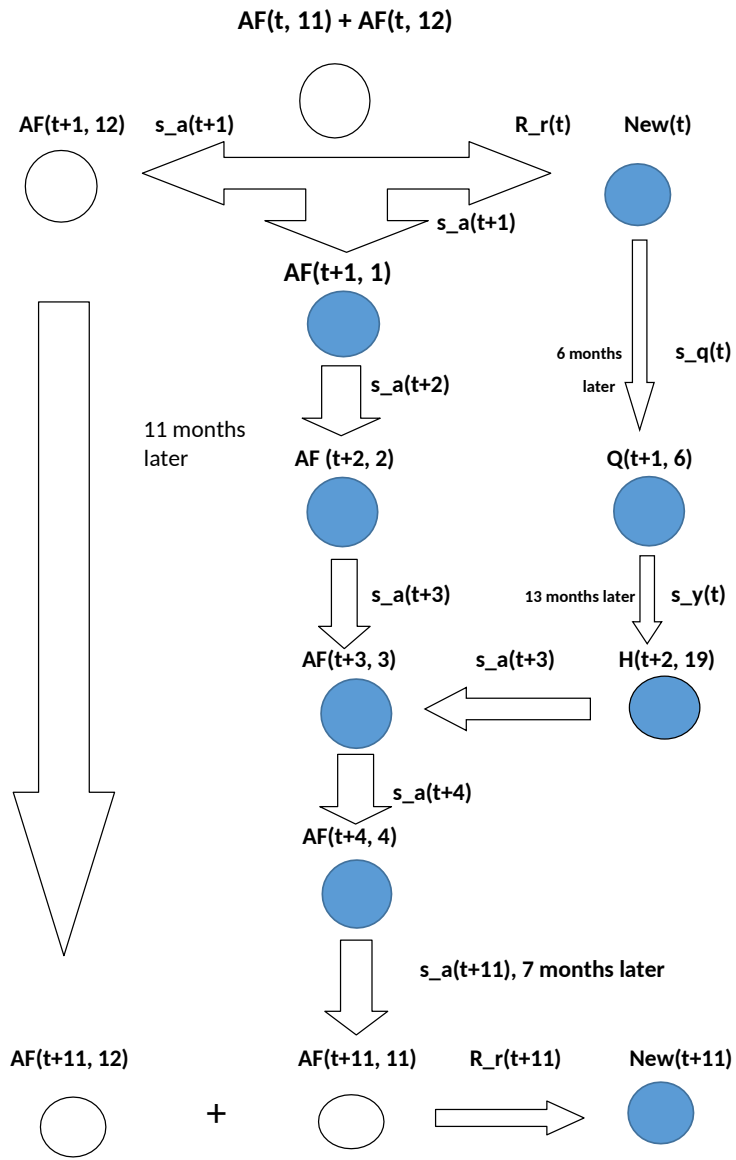}\\ \vspace{30mm}
\caption{Flow chart showing the reproduction and recruitment processes in topi.
The notations are defined in the text.}
\label{fig:flow_chart12}
\end{figure}

The adult males $AM(t)$ are not split into subgroups. 
Male and female half-yearlings at time $t-1$, denoted by $H(t-1, 19)$, 
that join the adult class at time $t$, are denoted by $NAM(t)$ and $NAF(t)$, 
respectively. We denote by $R_c(t)$ the probability that an individual graduating
from the half-yearling
class to the adult class is a female at time $t$ (henceforward referred to as sex ratio). $NAF(t)$ is therefore the number of new
adult female recruits that can conceive at time $t$ and 
give birth 8 months later. So, we add the new adult female recruits to
$AF(t, 3)$, the number of adult females that gave birth exactly three months
ago, and track future changes in the resulting total number. Note that 
$F(t) = \sum_{\ell=1}^{12}AF(t, \ell)$.
From the preceding definitions, it follows that a total of only 
$AF(t, 11)+AF(t, 12)$ females can conceive at time $t$. All these processes
are illustrated diagrammaticality in Figure \ref{fig:flow_chart12}.

\subsection{{\bf Stochastic topi population dynamics} \label{subsec:stoch_topi_pop_dyn}}

The stochastic topi population dynamics model for time $t = 2, \ldots, T$ can then be 
cast as in  Table \ref{tab:state_process}.
As defined before, $AF(t, 1)$ is the number of females that gave birth exactly
one month ago. We assume that the number of females that gave birth at 
time $t-1$ is the same as the number of newborns recorded at time $t-1$.
This is a slight underestimate because of unobservable calf mortalities.
The population dynamics of adult females is thus modelled as given in Table \ref{tab:state_process}.

\subsection{{\bf Initialising population size for topi age and sex classes} \label{subsec:initial_pop}}

The time $t$ = 0 for our data corresponds to June 1989, one month before the start of the 
ground sample surveys. To model population size at time $t=1$, we need to know the 
initial population size at time $t$ = 0. We denote the initial population distribution
by $New_0$ = Number of newborns at time $t$ = 0; $Q_0(k)$ = Number of quarters of age $k$ at time
$t$ = 0, $1\leq k \leq$6; $H_0(k)$ = Number of half-yearlings of age $k$ at time $t$ = 0, 
$1\leq k \leq$19; $AF_0(k)$ = Number of adult females that gave birth $k$ months before time $t$ = 0; 
$AM_0$ = Number of adult males at time $t$ = 0.

The initial states for the different age and sex classes at time $t$ = 0 
(namely, $AF_0(k)$, $1\leq k \leq 12$; $AM_0$, etc.) are 
assumed to follow normal distributions with means  
determined by the estimated population age and sex structure at the initial time $t$ = 0
and variances assumed to be all equal to 20,000.
The distributions of the rest of the initial states are given in Table \ref{tab:state_process}.

We used the aerial and ground survey data to estimate the 
unknown age and sex structure of the initial population. First, we selected the 
ground sample counts for the month of June. We calculated the proportion of 
animals in all the different age and sex classes in each of the 15 years spanning 1989 to 2003 and 
averaged the proportion for each age and sex class across all the 15 years. 
We used this average to represent the population proportion for each age and sex class at time
$t$ = 0 for the month of June. To derive the initial population size estimate for each age and
sex class, we multiplied the total population size estimated from the 
aerial survey  data at time $t$ = 0 (June 1989) with the average proportion for each age and 
sex class for June. We use these initial population size estimates for each age and sex class 
as the means of the normal distributions for the corresponding initial 
states, $New_0$, $AM_0$, etc.

\subsection{{\bf Integrating aerial survey data with demographic ground count data} \label{subsec:aerial_survey_int}}

We integrate the total population size estimates for the entire Masai Mara Ecosystem derived from the 
aerial surveys with the monthly demographic vehicle ground counts for the Masai Mara Reserve. This 
enables us to predict the total monthly population size estimates for the entire Masai Mara Ecosystem 
and for each of its four constituent zones. Let $B_t = New(t) + Q(t) + H(t) + F(t) + M(t)$ be the total topi population
size for the Mara Reserve and $T_t$ be the total topi population for all the four zones that collectively
constitute the Mara Ecosystem. We further assume that $T_t = K_t B_t$, where $K_t$ is the proportion $\frac{T_t}{B_t}$ at time $t$.
We integrate $T_t$ with the model for ground counts as follows.
\begin{eqnarray}
 \lambda_t^{T} &\sim& Gamma\left(\frac{(K_t B_t)^2}{\sigma_T^2}, \frac{K_t B_t}{\sigma_T}\right) \nonumber \\
 T_t &\sim& Poisson(\lambda_t^{T}) \label{eq:int_model}
\end{eqnarray}

\subsection{{\bf Observation process model} \label{subsec:observation_pro}}

The hidden states $(New(t), Q(t), H(t), F(t), M(t))$ are linked to the observed 
counts ($new(t),$ $q(t)$, $h(t)$, $f(t)$, $m(t)$) assuming Poisson distributions
as described in Table \ref{tab:state_process}.
We assume that $new(t)$ has expected value $New(t)$. But some newborn topi are almost certainly
missed during the ground surveys because topi hide their young for some time soon after birth, 
large carnivores kill some newborns whereas others may simply be missed due to visibility bias. 
To account for potential underestimation, we multiplied $new(t)$ with a correction factor of 1.7, based on 
experimentation, before determining birth rates in 
Section \ref{subsec:deter_surv_eqns}. More 
generally, however, sightability bias in $new(t)$ can be modeled by allowing $new(t)$ to follow a 
Poisson distribution with parameter $\frac{\lambda_1(t)}{h(t)}$, where $h(t)$ is a proportionality 
factor.

\section{{\bf Predicting expected number of animals in each age and sex class using 
simultaneous linear equations} \label{subsec:surv_eqns}}

We used an interdependent system of linear regression equations \citep{Theil1971} to estimate 
the expected total number of animals in each age and sex class present in the ground sample 
in each month for use in determining initial population size estimates. The current month endogenous variables appear as regressors in equations for 
other age or sex classes in the system of simultaneous equations. The model accounts for 
potential correlation of errors for the set of related regression equations to improve the
efficiency of parameter estimates. The modelling framework uses estimation 
procedures that produce consistent and asymptotically efficient estimates for the system of
linear regression equations. We imposed linear restrictions on some of the parameter estimates.

\subsection{{\bf Notations for predictors of birth recruitment and survival rates} \label{subsec:notations_cov}}

Denoting time by $t$ and the 174 months covered by the vehicle ground counts by $t = 1, \ldots, 174$,
we introduce a set of notations for the covariates used to model topi birth and adult recruitments, 
age-specific survival rates and adult sex ratio. The notations and their descriptions are 
summarized in Table \ref{tab:notations}.

\subsection{{\bf Determining the birth recruitment, sex ratio and age-specific survival 
functions} \label{subsec:deter_surv_eqns}}

The method used to determine the functional forms of the relationships between the birth and adult 
recruitments, sex ratio and age-specific survival rates and the various covariates is described in 
Web Appendix F.
The functional forms of the logit regression models relating birth and adult recruitments and 
age-specific survival rates and adult sex ratio to covariates are given by the equations 
in Table \ref{tab:logit_models}.

\subsection{{\bf Determining predation risk} \label{subsec:deter_risk_pred}}

The birth and survival rates are adjusted for environmental seasonality 
and seasonality in predation risk. During the dry season (July-October) migratory 
herbivores occupy the Mara, generating a superabundance of food for large 
predators thereby considerably reducing predation risk
for resident large herbivores, such as topi. So, we assume, based on parameter tuning,
that predation risk for topi during the dry season is 70\% of the risk during the wet 
season when the migrants are absent from the Mara. 
Also, adult male topi often fight each other and defend mating territories, potentially elevating
their susceptibility to predation. Thus, we assume, also based on parameter tuning, that the
survival rate for adult males is 99.7\% that of adult females.

\subsection{{\bf Prior distributions} \label{subsec:prior_distrib}}

The prior distributions for the regression coefficients 
$\bgamma^{S} = (\gamma^{S}_1, \ldots, \gamma^{S}_9)$,
$\bgamma^{R} = (\gamma^{R}_1, \ldots, \gamma^{R}_9)$, 
$\bgamma^{Q} = (\gamma^{Q}_1, \ldots, \gamma^{Q}_{14})$,
$\bgamma^{Y} = (\gamma^{Y}_1, \gamma^{Y}_2, \gamma^{Y}_3)$ and 
$\bgamma^{A} = (\gamma^{A}_1, \ldots, \gamma^{A}_9)$ of the logit regression models
are assumed to be normal with the same means  
and diagonal covariance matrices with diagonal elements equal to the estimated covariance 
components for the corresponding regression coefficient estimates for models in Table 3. 
The empirical choice of the priors ensures good mixing of the MCMC chains.

We assumed a prior on $\sigma^2$ in Section \ref{subsec:aerial_survey_int} 
following
a gamma distribution, Gamma(50,000, 0.000001), which has a very large average variance for 
topi population size of 50,000,000,000. Prior distributions for the initial states 
($New_0$, $Q_0(k)$, etc.) are specified in Section \ref{subsec:initial_pop}. The prior 
on $K_t$ is taken to be Beta(5402.23, 4182.9). This prior is determined by using the empirical proportions of
the total topi population size 
in each of the four zones constituting the Masai Mara Ecosystem derived from the aerial survey monitoring 
data for July 1989 to December 2003. The Dirichlet distribution is fitted to the data on proportions and Maximum 
Likelihood estimates of its parameters are obtained using the ``DirichletReg'' package in
R. The marginal beta 
distribution of the Dirichlet distribution is obtained as Beta(5402.23, 4182.9). $\sigma_T$ is fixed at
1906.42, the average standard deviation of the topi population size estimates for the 
Mara Ecosystem based on the aerial survey data for July 1989 to December 2003. 

\section{{\bf Full conditional distributions and convergence of MCMC chains} 
\label{sec:post_dist_converge_mcmc}}

The forms of the full conditional distributions of $New(t)$, $Q(t, k)$, $H(t, k)$, $NAF(t)$, 
$NAM(t)$ and other parameters
are presented in Web Appendix G. The functional 
forms of the full conditional distributions of $\bgamma^R$,
$\bgamma^Q$, $\bgamma^Y$ and $\bgamma^A$ (collectively refered to as $\bgamma$'s) 
and other parameters are also presented in Web Appendix G. 

The functional forms of the full conditional distributions for the 
$\bgamma$'s are not conformable to Gibbs sampling. Moreover, the resulting Metropolis-Hastings chain for 
the $\bgamma$'s converge quite slowly, making the algorithm highly inefficient. To 
accelerate the rate of convergence of the chain we implement 
Transformation Markov Chain Monte Carlo (TMCMC) at the Metropolis-Hastings step. A theoretical 
discussion of TMCMC can be found in \cite{DuttaBhattacharya2014}. Details on 
how TMCMC was specialized for our chain are discussed in Web Appendix H. The MCMC
simulations were continued for 1,500,000 iterations after discarding the initial 500,000 
iterations. The convergence of the MCMC chains for each parameter was assessed informally using trace plots.
All the trace plots show evidence of convergence and are provided in Web Figures 1 and 2.

\subsection{{\bf Model validation using balanced bootstrap 
sampling} \label{subsec:model_validation}}

To establish robustness of the model, we performed a model validation test using balanced
bootstrap samples. We first drew 10 different samples from each of the 
75 aerial surveys conducted between 1977 and 2018 using balanced bootstrap 
sampling. The balanced bootstrap selection was performed by using the algorithm 
of \cite{Gleason1988} in SAS PROC SURVEYSELECT. 
The balanced bootstrap method was used to  
select 10 samples from each of the total of 75 
aerial surveys with equal probability and with replacement, where each aerial 
survey had 232 to 705 sampling units each measuring 
5 km $\times$ 5 km, 2.5 km $\times$ 5 km  or 10 km $\times$ 5 km.
Because the bootstrap selection is
balanced the overall total number of selections is the same for each sampling 
unit \citep{Davison1986}. 
We then estimated the total topi population size for each bootstrap 
sample using Jolly's Method 2 \citep{JollyGM1969} for transects of unequal lengths.

\subsection{{\bf Cross-validation results} \label{subsec:cross-valid}}

Before running the model to produce the parameter estimates and interpreting 
their significance, we validated the model to establish its suitability for 
predicting population dynamics. The estimates of population sizes from the
bootstrap samples served as the actual population size of a hypothetical topi 
population. Using these estimates, we generated a set of 10 time series of 
hypothetical ground survey data each of length 174 months and having the same
age and sex classes as the actual ground sample count data 
(further details in Web Appendix I).
We call these hypothetical ground data sets generated data and the corresponding 
population size estimates bootstrap population estimates. We then fit our 
model separately to each of the generated data and obtained estimates of 
total population size and the corresponding 95\% credible limits of the estimates from 
the state process model in Section \ref{sec:des_model}. We call these 
estimates generated estimates. Next, we compare the bootstrap population 
estimates with the generated
estimates by observing whether the bootstrap estimates lie within the 
95\% credible limits of the generated estimates for each of the 10 bootstrap 
population time series. However, running the model separately for each 
time series is computationally very expensive. To reduce the computing time
we used the idea behind the Importance Resampling MCMC (IRMCMC)
proposed by \cite{BhattaHaslett2007}. Though developed for inverse problems, 
this method can be generalised to tackle the current problem. The precise 
details of this generalization are discussed in Web Appendix I.1. We ran our model using IRMCMC for each of the 
generated data sets (10 time series) and calculated the percentage coverage of the bootstrap 
populations at each of the 174 time points. The coverages for the 
10 time series of bootstrap samples vary from 85\% to 97\%. 
The bespoke R code written in R software version 4.1.1, 
used to implement the population dynamics model is provided in Web Appendices P
and Q, whereas the SAS  codes used to fit the simultaneous linear equations and relate birth 
recruitment and survival rates and adult sex ratio to the covariates  are provided in Web Appendix R.

\section{{\bf Results} \label{sec:results}}

\subsection{{\bf Topi population trajectory by age and sex class} \label{subsec:under_pop_dyn}}

\begin{figure}
  \centerline{\includegraphics[width=4.5in, angle=270]{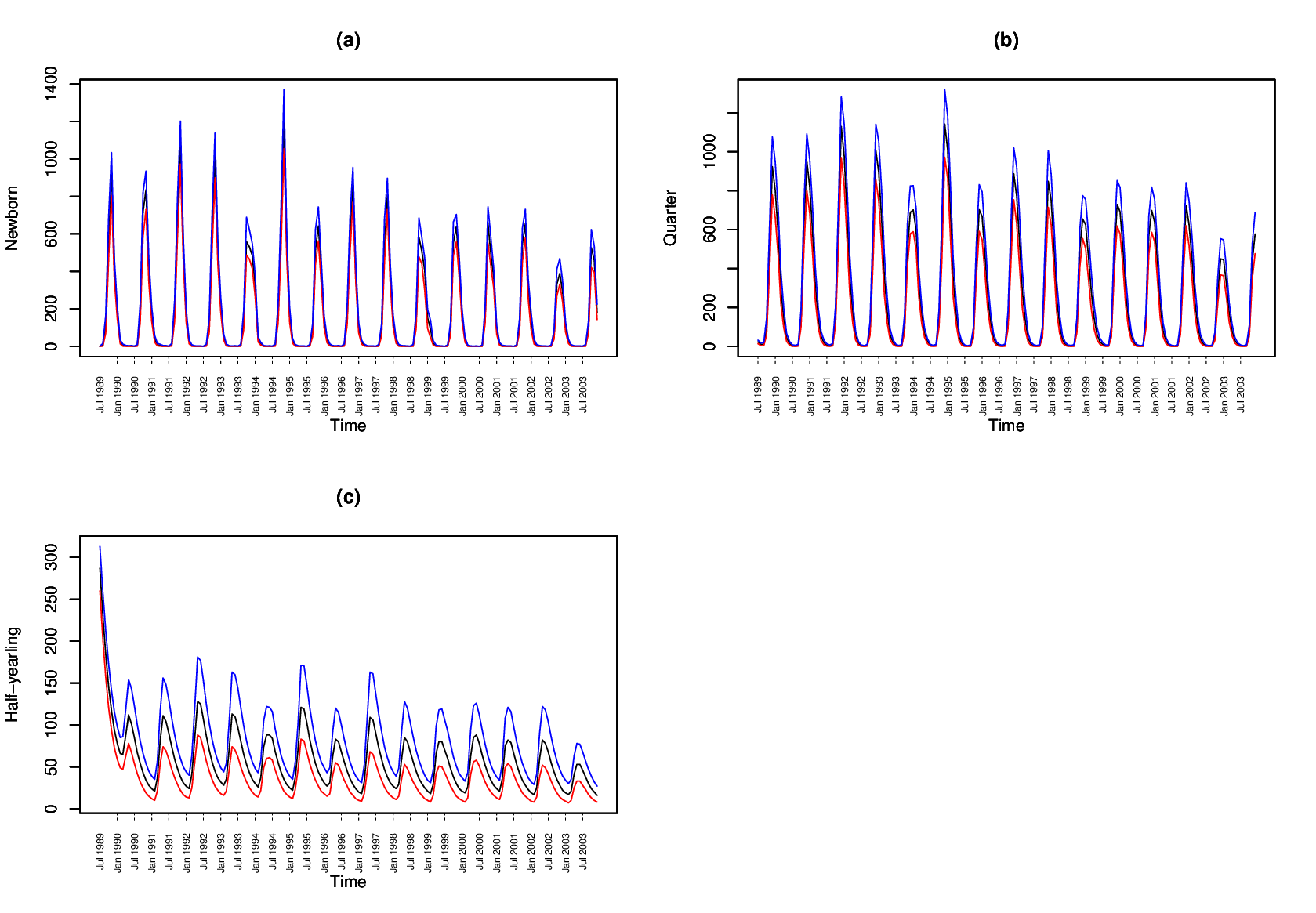}}
  \caption{Modelled (black lines) population trajectories and the 
        associated 95\% lower (red lines) and upper (blue lines) credible limits 
        for a) newborn, b) quarter and c) half-yearling topi.}
\label{fig:plot_newborn}
\end{figure}

\begin{figure}
  \centerline{\includegraphics[width=4.5in, angle=270]{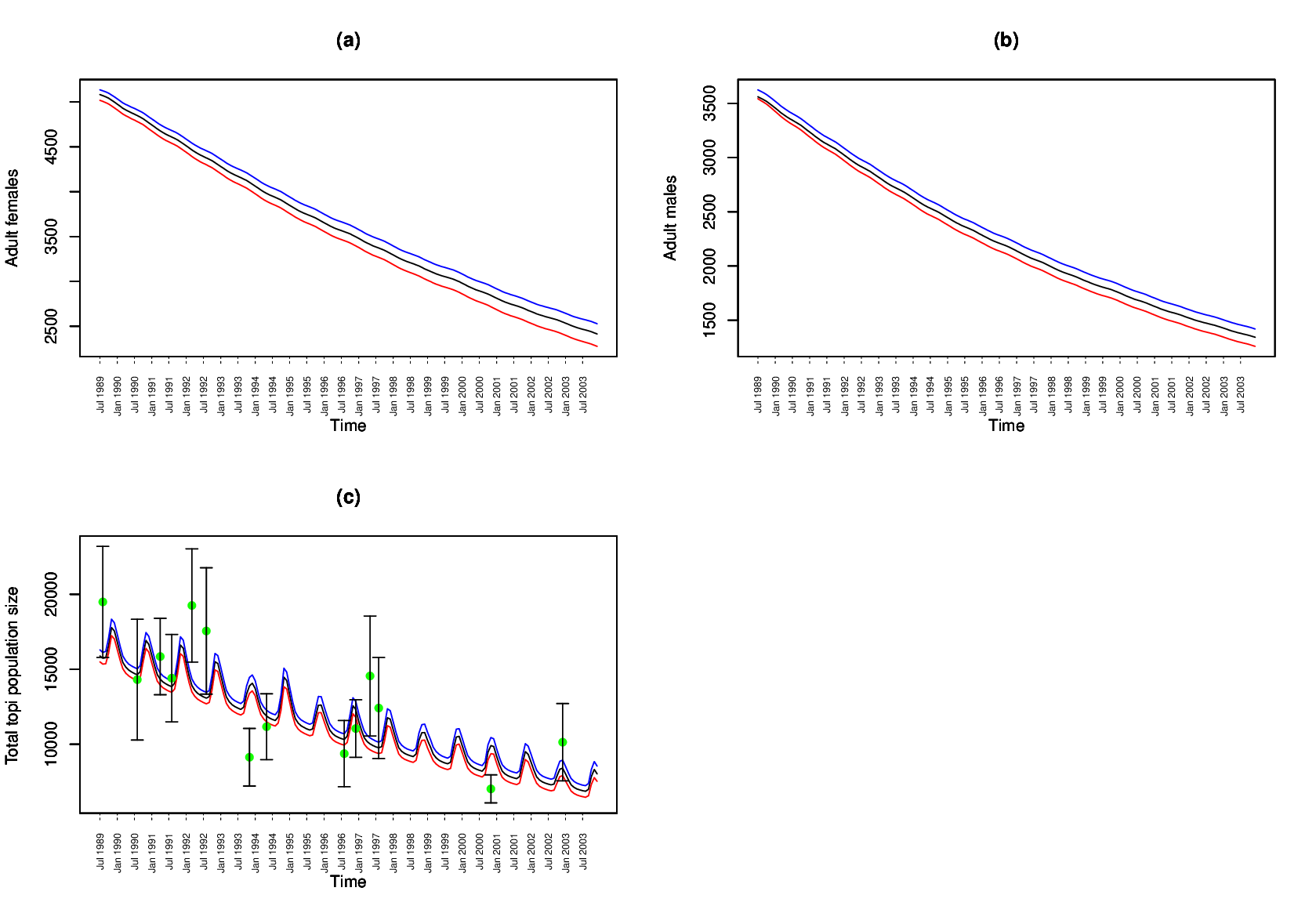}}
  \caption{Modeled (black lines) population trajectories and the 
        associated 95\% lower (red line) and upper (blue line) credible limits 
        for a) adult female, b) adult male and c) total topi population size 
        overlaid with the population size estimates (green filled circles) and their standard errors 
        (vertical whiskers) based on the DRSRS aerial surveys.
        The DRSRS population size estimates and their standard errors were multiplied
        by 1.3 to correct for sightability bias \citep{Stelfoxetal1986}.}
\label{fig:plot_adult}
\end{figure}

The model captures the essential and well-known features of the Mara topi 
population dynamics. First, it accurately captures the declining population trajectory of
all the age and sex classes; adult female, adult male, half-yearling, quarter-size and 
newborn topi in the Mara between 1989 and 2003 (Figures~\ref{fig:plot_newborn} and \ref{fig:plot_adult}). 
Second, it accurately tracks inter-annual variation in the reproductive seasonality, phenology,
synchrony and prolficity of topi births and juvenile recruitment 
(Figure~\ref{fig:plot_newborn}; \citeauthor{SinclairMduma2000}, \citeyear{SinclairMduma2000}; 
\citeauthor{OgutuPiepho2010}, \citeyear{OgutuPiepho2010}, \citeyear{Ogutupiepho2011}).
Figures \ref{fig:plot_newborn} and \ref{fig:plot_adult}
show the observed and predicted population sizes for each age and sex class 
for each of the 174 observation points (July 1989--December 2003) and the 
associated 95\% credible limits averaged across the 1,500,000 MCMC 
simulation replications. The 95\% credible 
intervals for the predicted values are not too wide 
indicating convergence of the MCMC chains. The birth recruitment rates 
(Figures \ref{fig:plot_newborn} and Web Figure 5a) show strong seasonality consistent with the pronounced 
reproductive seasonality characteristic of the Mara-Serengeti topi.
Further, the prolificacy of topi births is strongly time-varying, reflecting the controlling
influence of the seasonally and inter-annually varying rainfall 
(\citeauthor{OgutuPiepho2014}, \citeyear{OgutuPiepho2014}, 
\citeyear{Ogutuetal2015b}).
The pronounced seasonality in prolificacy of births and birth
recruitment also carry over to the trajectories of the quarter and 
half-yearling size 
classes (Figures 2b--c) but not to the adult age class (Figures 3a--b).
The persistent declines in the trajectories of topi birth recruitment, quarter and half-yearling classes, adult males and
females and the overall topi population size are consistent with the
overall topi population decline in the Mara Ecosystem from 1977 to 2018 
(Figure 3c; \citeauthor{Ogutupiephoetal2016}, \citeyear{Ogutupiephoetal2016}; 
\citeauthor{Veldhuis2019}, \citeyear{Veldhuis2019}). Finally, there is evident seasonality
in the overall topi population trajectory generated by the strong 
seasonality of births and juvenile recruitment in the ecosystem (Figure 3c). 

\subsection{{\bf Adult female recruitment and females available 
to conceive} \label{subsec:adult_female_recruit_con}}

Adult female recruitment is strongly seasonal, consistent with the 
seasonality in births and juvenile recruitment. The expected number of new 
adult females recruited into the population per month peaked in 1989--1990 but was noticeably lower in the other years (Web Figure 3a). Moreover, the
number of adult females that gave birth exactly 8 months ago (Web Figure 3b) and
the number of adult females that were available to conceive (Web Figure 3c) decreased 
persistently and markedly. The latter reduced from a maximum of nearly
4,000 in 1989 to around 2,000 animals by 2003 (Web Figure 3c). Thus, the decline 
in newborns was associated with a persistent decline in the number of adult females. Notably also, 
the severe 1993 drought resulted in evidently fewer adult females (in 1994) that gave birth in 1993
(Web Figure 3b) or that were available to conceive in 1994--1995 (Web Figure 3c).

\subsection{{\bf Temporal variation in age structure and adult sex ratio} \label{subsec:temp_var_age_struc}}

The expected population proportions of newborns, quarter-size and 
half-yearlings all tended to increase as the topi population 
decreased but the proportion of the adult population segment remained rather
constant over time apart from sustained seasonal oscillations from 1989 
to 2003 (Web Figures 4a--d).

\subsection{{\bf Factors influencing birth recruitment and survival rates and adult
sex ratio} \label{subsec:fact_influ_birth_rec}}

The posterior means, standard deviations and 95\% credible intervals 
for parameters of the models relating birth and adult recruitments, survival rates
and adult sex ratio to various covariates are summarized in Web Tables 4--8. Similarly, 
posterior densities for a sample of the parameters are displayed in Web Figures 7--10.
There was evident seasonality in birth recruitment and survival of 
quarter-size, half-yearling and adult topi (Web Figures 5a-d). 
Birth recruitment was negatively density-dependent for the declining topi population and was also influenced
by prior rainfall and average temperatures (Web Table 4). Besides seasonality,
quarter-size survival was influenced only by prior rainfall (Web Table 5). 
Half-yearling survival fluctuated seasonally and was apparently influenced only by past rainfall (Web Table 6). 
Lastly, adult survival was negatively density-dependent and also varied 
with past rainfall amounts (Web Table 7). Adult sex ratio varied seasonally and 
with rainfall, minimum and maximum temperatures but was apparently not density-dependent 
(Web Table 8).

\subsection{{\bf Birth recruitment rates, survival rates and sex ratio}
\label{subsec:birth_rec_surv_sex}}

The estimated birth recruitment, survival rates and sex ratios are consistent with 
expectation and are displayed in Web Figures 5--6. Topi birth recruitment 
was strongly seasonal, unusually low during the 1993-1994 drought  and the highest during 1994-1995 
with good rainfall (Web Figure 5a). The survival rates for quarter size (Web Figure 5b), half-yearling (Web Figure 5c) and adult 
(Web Figure 5d) topi age classes showed strong and sustained seasonality. Adult survival was the lowest
during 1999--2001, coincident with the extreme 1999--2000 drought (Web Figure 5d). The proportion of adult
females increased from a monthly maximum of 59\% to 64\% whereas the proportion of adult males 
decreased correspondingly from a peak of about 41\% to 35\% between 1989 and 2003 (Web Figures 6a--b).

\subsection{{\bf Comparing predicted topi population size with aerial survey data}
\label{subsec:compare_aerial_survey}}

Fourteen of the 75 DRSRS aerial surveys for the Masai Mara Ecosystem for 
1977--2018 (Section \ref{aerial_sample_data}) fell within the period spanned by 
the ground survey data (July 1989--December 2003). The total population size estimates from these 
14 aerial surveys were in reasonable agreement with the total topi population 
size predicted by the Bayesian state-space model. Notably, the estimated total topi population size 
was within the 95\% confidence limits of most of the  total population size estimates derived from 
the 14 DRSRS aerial surveys (Figure \ref{fig:plot_adult}). Moreover, trends in 
the estimated topi population size for each of the four zones constituting the Mara 
ecosystem (Web Figures 11a--d) are also in good agreement with 
the corresponding aerial total population size estimates and show that topi numbers worryingly declined 
persistently throughout the Mara Ecosystem regardless of the degree of protection -- highest in the Mara Reserve,
intermediate in the conservancies and lowest in the unprotected agro-pastoral lands of Siana and the Loita
Plains. However, topi were substantially  more abundant in the protected reserve and semi-protected conservancies than 
in the unprotected agro-pastoral lands.

\section{{\bf Discussion} \label{sec:discussion}}

We develop a flexible and general, integrated state-space model 
in a Bayesian framework for estimating large herbivore population 
demographic parameters, dynamics and the associated uncertainties. The model is 
illustrated using 
the Mara-Serengeti topi population inhabiting the World-famous Greater Mara-Serengeti Ecosystem 
of Kenya and Tanzania in East Africa. The state-space model allows 
estimation of both process and observation error 
variances (\citeauthor{ValpineHastings2002}, \citeyear{ValpineHastings2002};
\citeauthor{Bucklandetal2004}, \citeyear{Bucklandetal2004}).
The model incorporates age and sex structure and key life-history 
characteristics (e.g., gestation length, lactation period, pregnancy status,
females available to conceive, one young per birth event) and life-history strategies 
(e.g. feeding style, grazer for topi, and foraging style, resident for topi) essential to understanding large herbivore 
population dynamics and efficiently 
integrates ground demographic survey with aerial 
survey data. We relate birth and adult recruitments and age-class specific 
survival rates and adult sex ratio to various covariates, such as prior 
population density, predation risk, past rainfall and temperature, 
environmental seasonality and their 
interactions. 
To estimate model parameters, we used the 
MCMC method in a Bayesian framework because of its flexibility
(\citeauthor{Brooksetal2004}, \citeyear{Brooksetal2004}; 
\citeauthor{HoyleMaunder2004}, \citeyear{HoyleMaunder2004}; 
\citeauthor{Schaubetal2007}, \citeyear{Schaubetal2007}; 
\citeauthor{Finke2019}, \citeyear{Finke2019}). 
The convergence of the MCMC chains of the parameters is accelerated using the 
Transformation MCMC (TMCMC) technique \citep{DuttaBhattacharya2014} and 
the idea behind IRMCMC \citep{BhattaHaslett2007} to reduce the 
computational time for model validation. 

The predicted population trajectories show persistent and marked declines in the Mara 
topi population from 1989 to 2003
in accord with the trends derived from aerial surveys for
1977--2018 (\citeauthor{Ogutupiephoetal2016}, 
\citeyear{Ogutupiephoetal2016}; \citeauthor{Veldhuis2019},
\citeyear{Veldhuis2019}).
Importantly, whereas birth and adult recruitments fluctuated around a stable average,
the survival rates for quarter size,
half-yearling size and adults decreased gradually and contemporaneously 
with the overall topi population decline. The disturbing and sustained 
decline in numbers of topi and other species in the Mara and across Kenya
\citep{Ogutupiephoetal2016} and continental Africa \citep{Craigieetal2010} increases
the urgency for establishing their leading causes and developing 
effective population conservation and recovery strategies.

The modelling framework can be extended to (i) identify the most
influential processes driving population declines and assess their relative importance,
(ii) test interesting ecological hypotheses, (iii)
enable rigorous forward-projection of large herbivore population dynamics 
allowing for both parameter and future demographic uncertainty 
\citep{HoyleMaunder2004}. Further extension can also (iv) enable assessment of likely future populating
trajectories under
various scenarios of climate change, land use change, socio-economic development,
human population growth, livestock population density, conservation and 
management interventions, such as formation of new wildlife conservancies 
(\citeauthor{Rhodesetal2011}, \citeyear{Rhodesetal2011};
\citeauthor{Maunderetal2015}, \citeyear{Maunderetal2015};
\citeauthor{Mosnieretal2015}, \citeyear{Mosnieretal2015}). Moreover, the monthly
time step allows the model outputs to be used to study potential shifts in 
reproductive seasonality, phenology, synchrony and prolificacy of births,
juvenile recruitment and adult sex ratio 
(\citeauthor{OgutuPiepho2010}, \citeyear{OgutuPiepho2010}; 
\citeyear{Ogutupiepho2011}, \citeyear{OgutuPiepho2014}, \citeyear{Ogutuetal2015b})
in response to future changes in climate and other factors.
It would also be interesting and useful to adapt the model for 
other species, with contrasting life-history traits and strategies as well as trophic relations, 
such as hartebeest ({\it Alcelaphus buselaphus cokeii}), 
impala ({\it Aepyceros melampus}), waterbuck ({\it Kobus ellpsyprimnus}),
zebra, warthog ({\it Pharcocoerus africana}) and giraffe 
({\it Giraffa camelopardalis}).

The model can also be extended to include several 
additional features. First, females that are reproducing for the
first time can be allowed to have a lower pregnancy rate (86\%) than older (100\%)
females \citep{Duncanthesis1975}. Second, females that lose their calves soon after
birth can be moved to the group of females available to conceive. 
Third, the survival rate of calves aged 0 to 1 month can be explicitly
incorporated in the model, especially if empirical estimates of such rates can be 
obtained. Fourth, the dependence of birth recruitment and 
age-specific survival rates and sex ratio on covariates may alternatively be
made time-varying to potentially better account for temporal 
variation in the influence of the covariates. Fifth, the model can be made spatially 
explicit to allow for spatial variation in the type and intensity of 
factors influencing survival and recruitment rates and sex ratio. 
Sixth, sightability bias in newborns,  $new(t)$, can be modeled explicitly by 
specifying $new(t)$ to follow a Poisson distribution with parameter $\lambda_1(t)/h(t)$, 
where the proportionality 
constant $h(t)$ is sampled from a suitable probability distribution.
Seventh, birth recruitment and survival rates and sex ratio can be related to additional
covariates, particularly
anthropogenic covariates, such as human population
density, livestock population density, settlement density and progressive habitat loss.
Lastly, birth recruitment, survival rates and sex ratio can all be related to one set of covariates collected
at the monthly time scale using the ground demographic data whereas the inter-annual variation in the 
overall population growth rate can be related to the other set of covariates collected at the annual time 
scale using the aerial survey data.

\section*{{\bf Acknowledgements}}
The ground surveys were funded by the World Wide Fund for Nature--East Africa Program 
(WWF--EARPO) and Friends of Conservation (FOC) and the aerial surveys by the Kenya Government through the DRSRS.  
This project has received funding from the European Union's Horizon 2020 research and
innovation programme under Grant Agreement No. 641918. JOO was additionally supported
by a grant from the German Research Foundation (DFG; Grant \# 257734638).
Further information on acknowledgment 
(data generation and provision, permissions, additional financial and logistical support, etc)
is provided in Web Appendix O.


\bibliographystyle{biom} 
\bibliography{biomsample_bib}

\section*{Supporting Information}

Web Appendices, Tables, and Figures referenced in Sections 2--6 are provided 
in the files Web Appendices A-O.pdf. The R and SAS codes for the
method and the simulation are provided in Web Appendix Q.R, Web Appendix Q.P and 
Web Appendix R.sas. The ground survey dataset is available in TopiData.csv 
whereas the aerial survey dataset is provided in AerialData.csv.

\clearpage

\newpage

\begin{table}
\begin{center}
\caption{State-space and observation process model formulations. All the notations are explained in 
Sections \ref{sec:des_model}, \ref{subsec:split_age_class},
\ref{subsec:stoch_topi_pop_dyn}, \ref{subsec:observation_pro} and \ref{subsec:initial_pop}
in the text.}
\small
\begingroup
\setlength{\tabcolsep}{10pt} 
\renewcommand{\arraystretch}{1.5} 
\begin{tabular}{|l|p{.8\textwidth}|}
\hline     
Model & Distributions \\ \hline \\
State process   &  $New(t) \sim Binomial(AF(t, 11)+AF(t, 12), R_r(t))$   \\  
for newborns,    &  $Q(t, 1) \sim Binomial(New(t-1), s_q(t))$  \\  
quarter size, half-yearlings, &  $Q(t, k) \sim Binomial(Q(t-1, k-1), s_q(t)), k = 3, \ldots, 6$   \\  
adult females and adult males &  $H(t, 1) \sim Binomial(Q(t-1, 6), s_h(t))$  \\ 
for time $t = 2, \ldots, T$  &  $H(t, k) \sim Binomial(H(t-1, k-1), s_h(t)), k = 7, \ldots, 19$ \\
                &   $(NAF(t), NAM(t)) \sim Multinomial(H(t-1, 19), R_c(t)\times s_a(t), (1-R_c(t))\times s_a(t))$ \\
                &  $AM(t) \sim Binomial(AM(t-1)+NAM(t-1), s_a(t))$ \\ 
                \vspace{3mm} \\
State process   &  $AF(t, 1) \sim Binomial(New(t-1), s_a(t))$ \\
for breeding adult females &  $AF(t, k) \sim Binomial(AF(t-1, k-1), s_a(t)), k = 2, 3$ \\
for time $t = 2, \ldots, T$ &  $AF(t, 4) \sim Binomial(AF(t-1, 3)+NAF(t-1), s_a(t))$ \\
                &  $AF(t, k) \sim Binomial(AF(t-1, k-1), s_a(t)), k = 5, \ldots, 11$ \\
                &  $AF(t, 12) \sim Binomial(AF(t-1, 11)+AF(t-1, 12)-New(t-1), s_a(t))$ \\ 
                \vspace{3mm} \\
Initial state process & $New_0 \sim Binomial(AF_0(11)+AF_0(12), R_r(t))$   \\  
at time $t$ = 0       & $Q(1, 1) \sim Binomial(New_0, s_q(1))$  \\  
                          & $Q(1, k) \sim Binomial(Q_0(k-1), s_q(1)), k = 3, \ldots, 6$   \\  
                          & $H(1, 1) \sim Binomial(Q_0(6), s_h(1))$  \\ 
                          & $H(1, k) \sim Binomial(H_0(k-1), s_h(1)), k = 7, \ldots, 19$ \\
                          & $(NAF(1), NAM(1)) \sim Multinomial(H_0(19), R_c(1)\times s_a(t), (1-R_c(1))\times s_a(t))$ \\
                          & $AM(1) \sim Binomial(AM_0, s_a(1))$ \\ 
                          & $New(1) \sim Binomial(AF(1, 11)+AF(1, 12), R_r(t))$ \\
                          & $AF(1, 1) \sim Binomial(New_0, s_a(1))$ \\
                          & $AF(1, k) \sim Binomial(AF_0(k-1), s_a(1)), k = 2, 3, \ldots, 11$ \\
                          & $AF(1, 12) \sim Binomial(AF_0(11)+AF_0(12)-New_0, s_a(1))$
                          \vspace{4mm} \\ 
Observation process       & $new(t) \sim Poisson(\lambda_{1}(t))$ \\
                          & $q(t) \sim Poisson(\lambda_{2}(t))$ \\
                          & $h(t) \sim Poisson(\lambda_{3}(t))$     \\  
                          & $f(t) \sim Poisson(\lambda_{4}(t))$  \\
                          & $m(t) \sim Poisson(\lambda_{5}(t))$, \\ 
                          &  with parameters $\lambda_{i}$, $i = 1, \ldots, 5$, each assumed to follow a gamma distribution 
as $Gamma(\alpha_{i}(t), \beta_{i}(t))$, $i = 1, \ldots, 5$,
where $\alpha_{1}(t) = \frac{New^2(t)}{\sigma^2}$ and $\beta_{1}(t) = \frac{New(t)}{\sigma^2}$. The parameter $\sigma^2$ is a constant and can be 
interpreted as the variance
of the actual population size. Similarly, we define $\alpha_{2}(t) = \frac{Q^2(t)}{\sigma^2}$, 
$\beta_{2}(t) = \frac{Q(t)}{\sigma^2}$, $\alpha_{3}(t) = \frac{H^2(t)}{\sigma^2}$, 
$\beta_{3}(t) = \frac{H(t)}{\sigma^2}$, $\alpha_{4}(t) = \frac{F^2(t)}{\sigma^2}$, 
$\beta_{4}(t) = \frac{F(t)}{\sigma^2}$, $\alpha_{5}(t) = \frac{M^2(t)}{\sigma^2}$
and $\beta_{5}(t) = \frac{M(t)}{\sigma^2}$. \\ \hline
                               
\end{tabular}
\endgroup
\label{tab:state_process}
\end{center}
\end{table}

\begin{table}
\begin{center}
\caption{ Notations for the covariates used for building the logit models
in Table \ref{tab:logit_models} and their descriptions.}
\small
\begingroup
\begin{tabular}{|l|p{.8\textwidth}|}
\hline 
Notation for covariate & Description of notation for covariate \\ \hline \\
$month(t)$ =     &  Calender month corresponding to January-December and numbered as 1, 2, $\ldots$, 12 \\ \hline 
$rain(7-11, t)$ = & Average total monthly rainfall including lags 6 to 10 at time 
$t$ (i.e., 7 to 11 months before the birth month)\\ \hline 
$Npop(t)$ =      &  Total topi population size at lag 7 at time $t$ 
(i.e., around conception time 8 months ago)  \\ \hline 
$mintemp(t)$ =  & Minimum temperature at time $t$ \\ \hline
$maxtemp(t)$ =  & Maximum temperature at time $t$ \\ \hline
$lagmin(\ell, t)$ = & Minimum temperature at lag $\ell$ at time
$t$, $\ell = 1, 2, \ldots$, 11 (i.e., up to 4 months pre-conception) \\ \hline
$lagmax(\ell, t)$ =  & Maximum temperature at lag $\ell$ at time $t$, $\ell = 1, 2, \ldots$, 11 \\ \hline
$Apop(t)$ =  &  Total population size at lag 1 at time $t$  \\ \hline
$lagrain(\ell, t)$ =  &  Total monthly rainfall at lag $\ell$ at time $t$, 
$\ell = 0, 1, 2, \ldots, 11$. Note that $lagrain(0, t)$ stands for total monthly
rainfall at time $t$  \\ \hline
$wet1(t)$ =  &  Total wet season rainfall at lag 1 (i.e., in the immediately preceding year) at time $t$ \\ \hline
$earlywet1(t)$ =  &  Total early wet season (November--February) rainfall at lag 1 (i.e., 
in the immediately preceding year) at time $t$ \\ \hline
$dry1(t)$ =  &  Total dry season rainfall at lag 1 (i.e., in the immediately preceding year) at time $t$ \\ \hline
$mavrain(\ell-p, t)$ =  & Moving average of rainfall between lags $\ell$ and $p$ ($\ell < p$) at time $t$ \\ \hline

\end{tabular}
\endgroup
\label{tab:notations}
\end{center}
\end{table}

\begin{table}
\begin{center}
\caption{Formulations of the relationships between the logits of recruitment and survival rates 
and covariates. All the notations are explained in Sections \ref{sec:des_model}, \ref{subsec:split_age_class},
\ref{subsec:stoch_topi_pop_dyn}, \ref{subsec:initial_pop} and \ref{subsec:notations_cov}
and in Table \ref{tab:notations} in the text. The covariates are explained  in Web Table 3. The models were fitted using 
the SAS Procedure GLIMMIX (Web Appendix R).}
\small
\begingroup
\setlength{\tabcolsep}{10pt} 
\renewcommand{\arraystretch}{3.5} 
\begin{tabular}{|l|p{.8\textwidth}|}
\hline     
Rates & Equation \\ \hline \\
Birth recruitment  &  $\log\left(\frac{R_{r}(t)}{1-R_{r}(t)}\right) = \gamma^{R}_1+\gamma^{R}_2 month(t)+\gamma^{R}_3 month^2(t)+
\gamma^{R}_4 month^3(t)+  
\gamma^{R}_5 rain(7-11, t)+\gamma^{R}_6 rain^2(7-11, t)+$
$\gamma^{R}_7 Npop(t)+\gamma^{R}_8 mintemp(t)+ \gamma^{R}_9 maxtemp(t)$ \\ 

Quarter size survival  &  $\log\left(\frac{s_{q}(t)}{1-s_{q}(t)}\right) = \sum_{k=1}^{12}\gamma^{Q}_k \delta_{k}(t)+
\gamma^{Q}_{13} dry1(t)+\gamma^{Q}_{14} mavrain(3-4, t)$ \\ 

Half-yearling survival  &  $\log\left(\frac{s_{y}(t)}{1-s_{y}(t)}\right) = \sum_{k=1}^{12}\gamma^{Y}_k \delta_{k}(t)+
\gamma^{Y}_{13} earlywet1(t)$ \\

Adult survival  &  $\log\left(\frac{s_{a}(t)}{1-s_{a}(t)}\right) = \sum_{k=1}^{2}\gamma^{A}_k \delta_{k}(t)+\gamma^{A}_3 Apop(t)+
\gamma^{A}_4 lagrain(4, t)+ 
\gamma^{A}_5 lagrain(5, t)+\gamma^{A}_6 lagrain(6, t)+\gamma^{A}_7 lagrain(7, t)+ 
\gamma^{A}_8 wet1(t)$ \\

Adult recruitment  &  $\log\left(\frac{R_{c}(t)}{1-R_{c}(t)}\right) = \sum_{k=1}^{2}\gamma^{S}_k \delta_{k}(t)+\gamma^{S}_3 wet1(t)+
\gamma^{S}_4 dry1(t)+  
\gamma^{S}_5 lagrain(0, t)+\gamma^{S}_6 rain(7-11, t)+\gamma^{S}_7 mintemp(t)+ 
\gamma^{S}_8 lagmin(2, t)+\gamma^{S}_9 lagmax(1, t)$ \\ \hline 
\end{tabular}
\endgroup
\label{tab:logit_models}
\end{center}
\end{table}

\end{document}